\documentclass[10pt,a4paper]{article}

\usepackage{ifthen}
\newboolean{acm}
\setboolean{acm}{false}

\usepackage{amssymb}
\usepackage{amsmath}
\usepackage{algorithmic}
\usepackage{algorithm}
\usepackage{color}
\usepackage{epsfig}
\usepackage{url}

\ifthenelse{\boolean{acm}}{}{
\newenvironment{proof}{\noindent{\bf Proof.}\hspace{0.05in}}{\par\mbox{}\par}
\newcommand{\qed}{\mbox{$\blacksquare$}}
}

\newcommand{\NN}{\mbox{$\mathbb{ N}$}}
\newtheorem{definition}{Definition}
\newtheorem{theorem}{Theorem}
\newtheorem{example}{Example}
\newtheorem{lemma}{Lemma}
\newcommand{\RR}{\mathbb{R}}

\newcommand{\SNotify}[1]{\mbox{Notify}\ifthenelse{\equal{#1}\empty}{}{(#1)}}

\newcommand{\SIn}[1]{\underline{#1}}
\newcommand{\SOut}[1]{\overline{#1}}
\newcommand{\STrue}{\mathbf{T}}
\newcommand{\SFalse}{\mathbf{F}}

\newcommand{\SSemL}{[\![}
\newcommand{\SSemR}{]\!]}
\newcommand{\SSem}[1]{\SSemL #1\SSemR}
\newcommand{\SSemD}[1]{\SSemL #1\SSemR^{\circ}}

\newcommand{\SIU}{\uplus}
\newcommand{\STI}{\!\in\!}
\newcommand{\SNarCP}[1]{\SOut{#1}}
\newcommand{\SNarC}{\SNarCP{N}}
\newcommand{\SNarI}{\SIn{N}}
\newcommand{\SNar}{\SOut{N}}
\newcommand{\SSInt}{\mathbb{I}}

\newcommand{\SFix}{\mbox{fix}}

\newcommand{\SNeg}[1]{\Vec{\neg}#1}
\newcommand{\SAssVar}[3]{#1\frac{#2}{#3}}

\newcommand{\SLift}[1]{\widehat{#1}}

\title{Efficient Solving of\\Quantified Inequality Constraints\\over the Real Numbers\footnote{This is a
    revised and extended version of an earlier paper~{\cite{Ratschan:01c}. Please note the
    changed terminology that should contribute to a wider accessibility of the results.}}}
\ifthenelse{\boolean{acm}}{
\author{STEFAN RATSCHAN\\
Max-Planck-Institut f{\"u}r Informatik, Saarbr{\"u}cken, Germany}}
{\author{Stefan Ratschan\\\url{Stefan.Ratschan@mpi-sb.mpg.de}}}

\ifthenelse{\boolean{acm}}{}{
\begin{document}
\maketitle
}

\begin{abstract}
  Let a quantified inequality constraint over the reals be a formula in the first-order
  predicate language over the structure of the real numbers, where the allowed predicate
  symbols are $\leq$ and $<$. Solving such constraints is an undecidable problem when
  allowing function symbols such $\sin$ or $\cos$. In the paper we give an algorithm that
  terminates with a solution for all, except for very special, pathological inputs. We ensure the practical
  efficiency of this algorithm by employing constraint programming techniques.
\end{abstract}

\ifthenelse{\boolean{acm}}{

\begin{document}

\maketitle
}{}

\section{Introduction}

The problem of solving quantified constraints over the reals has numerous applications (we
have created a web-page that lists more than fifty references~\cite{Ratschan:01e}).
However, it is undecidable, when allowing function symbols such as $\sin$ or
$\cos$~\cite{Tarski:51}, and highly complex when restricting oneselves
to addition and multiplication~\cite{Davenport:88,Weispfenning:88}.

In this paper we give an algorithm that nevertheless always terminates for inputs that are
stable in the sense that their truth value (in the case where all variables are
quantified) does not change under small perturbations of the occurring constants. For example,
the constraint $\forall x\; x^2+1\geq 0$ is stable, whereas $\forall x\; x^2\geq 0$ is not. 

Furthermore, we ensure the practical efficiency of the algorithm by basing it on
techniques from the field of constraint
programming~\cite{Benhamou:94a,Benhamou:97a,Sam-Haroud:96,Hentenryck:97}.  The basic idea
of these techniques is to reduce the average run-time of algorithms for computationally
hard problems by replacing expensive exhaustive search as much as possible by methods for
pruning elements from the search space for which it is easy to show that they do not
contain solutions.

In this paper we extend this idea to quantified inequality constraints for which all (free
and bound) variables are bounded to a closed interval: We try to prune elements from these
bounds for which it is easy to show that they do not contain solutions.  When we cannot
easily prune more elements, we do branching by splitting a bound into pieces (for
quantified variables this means replacing sub-constraints of the form $\forall x\STI
I\;\phi$ by $\forall x\STI I_{1}\;\phi\;\wedge\;\forall x\STI I_{2}\;\phi$ where
$I=I_{1}\cup I_{2}$, or the corresponding existential case). This gives us new
possibilities for pruning.  We repeat the two steps until we have pruned all elements (or
disproved the constraint).  For computing elements of the bounds that do contain
solutions we take the negation of the input constraint and again apply the above branch-and-prune
approach.

In the paper we formalize this approach, study its properties in detail, improve it for an
implementation, and do timings that show its efficiency. 


As a side-effect, this paper even improves the current methods for numerical constraint
satisfaction problems in the case where the solution set does not consist of finitely
many, isolated solutions, which---up to now---was essential for their efficiency. For
example, the book describing the system Numerica~\cite{Hentenryck:97b} explicitly states
that for inputs not fulfilling that property the method creates a huge number of boxes.


In order to be able to reuse existing theory, algorithms and software for solving atomic
inequality constraints (i.e., constraints of the form $t\geq 0$ or $t>0$, where $t$ is a
term), and to be able to benefit from further progress in this area, the paper employs a
parametric approach: It takes as input theory and algorithms from constraint programming,
and provides as output corresponding new theory and algorithms for solving quantified
constraints.  More specifically, building upon the notion of a narrowing
operator~\cite{Hentenryck:97,Benhamou:99,Hong:94b}, it takes as input: a specification
describing a consistency notion for atomic inequality constraints (e.g.,
box-consistency~\cite{Benhamou:94a}), and a narrowing operator that implements this
specification. It provides as output: a specification describing a corresponding
consistency notion for quantified constraints, a narrowing operator that implements this
specification, and an algorithm for computing approximate solutions of quantified
constraints over the reals that uses this narrowing operator for pruning. These outputs
are accompanied with proofs of their usefulness/optimality.

For the special case where the only allowed function symbols are addition and
multiplication, up to recently, all algorithms have been based on computer algebra
methods~\cite{Collins:75,Caviness:98}, which resulted in certain drawbacks (e.g., low
practical efficiency, restriction to polynomials, unwieldy output expressions).  In an
earlier paper~\cite{Ratschan:01a} the author of this paper proposed a scheme for solving
quantified constraints approximately that followed the idea of quantifier elimination by
cylindrical algebraic decomposition~\cite{Collins:75,Caviness:98}, but decomposed space
into floating-point boxes instead of semi-algebraic cells.
This approach was successful in showing that one can efficiently compute approximate
solutions of quantified constraints using interval methods. However it still had several
drawbacks. Especially, it was not clear when and how to optimize box splitting, because
the algorithm was not separated into (inherently exponential) search, and pruning. The
current paper provides a solution to this, and other, problems of the older approach.



The following special cases of the general problem have been studied using interval or
constraint satisfaction methods:

\begin{itemize}
\item The case of expressions of the form $\forall \vec{p}\;\phi(\vec{p},\vec{q})$, where
  $\phi(\vec{p},\vec{q})$ is a system of strict inequalities~\cite{Jaulin:96,Malan:97,Jaulin:01b}
  using methods that repeatedly bisect the free-variable space, and test after each
  bisection, whether the system of inequalities holds everywhere on the resulting box.
\item The case of expressions of the form $\forall p\;\phi(p,\vec{q})$, where
  $\phi(p,\vec{q})$ is a system of strict inequalities~\cite{Benhamou:00}, using methods
  that correspond to our case for universal quantification, conjunction and atomic
  constraints, but without branching in the universally quantified variables.
\item The case of quantified systems of equations where certain variables occur only once,
  and the quantifiers obey certain orderings (various results by S.~Shary, see just for
  example~\cite{Shary:02}).
\item The case of disjunctive constraints. This has been done in the discrete
  case~\cite{Jourdan:93,Hentenryck:95a}, and in the continuous case for disjunctive
  constraints occuring in interactive graphical applications~\cite{Marriott:01}, and in a
  similar way as in this paper for speeding up solving of factorizable
  constraints~\cite{Gransvillier:98}.
\end{itemize}

See also an overview on methods for solving quantified inequalities in
control~\cite{Dorato:00}. For improving box splitting strategies for inequality
constraints Silaghi, Sam-Haround and Faltings~\cite{Silaghi:01} use information from the
negation of the input constraints. A similar problem is the problem of extending the
bounds of universal quantifiers, for which a method based on constraint
propagation~\cite{Collavizza:99a} is provided for the special case of a system of
inequalities for which all variables are universally quantified.
Also for discrete domains there is a lot of recent
interest solving constraints with quantifiers~\cite{Bordeaux:02,Fargier:96}, or 
related stochastic constraints~\cite{Walsh:02}.

Some of the above~\cite{Collavizza:99a,Benhamou:00,Silaghi:01} take a similar approach of
using the negation of the input to compute positive information. However, they do not
address the question of being able to compute answers for all except unstable inputs.


The content of the paper is as follows: Section~\ref{sec:preliminaries} gives various
preliminaries; Section~\ref{sec:first-order-narr} introduces a framework for reusable
pruning based on the notion of narrowing operator; Section~\ref{sec:first-order-cons}
describes an according notion of consistency for quantified constraint that allows us to
specify the pruning power of narrowing operators; Section~\ref{sec:narrowing-algorithm}
gives a generic algorithm for pruning that implements a narrowing operator;
Section~\ref{sec:solver} bases an according branch-and-prune solver for quantified
constraints on this pruning algorithm.  Section~\ref{sec:effic-impl} discusses how to
arrive at an efficient implementation of such a solver; Section~\ref{sec:timings} presents
timings of such an implementation;  Section~\ref{sec:relat-class-algor} discusses the
relation of the results to symbolic quantifier elimination algorithms; and
Section~\ref{sec:conclusion} concludes the paper.

\section{Preliminaries}
\label{sec:preliminaries}

We fix a set $V$ of variables. A quantified constraint (or short: constraint) is a formula
in the first-order predicate language over the reals with predicate and function symbols
interpreted as suitable relations and functions, and with variables in $V$. We take over a
large part of the according predicate-logical terminology without explicit definitions.

In this paper we restrict ourselves to the predicate symbols $<$, $>$, $\leq$, and $\geq$,
and assume that equalities are expressed by inequalities on the residual (i.e., $f = 0$ as
$| f | \leq\varepsilon$ or $f^2\leq\varepsilon$, where $\varepsilon$ is a small positive
real constant\footnote{The constant $\varepsilon$ needs to be non-zero because otherwise
  solutions would vanish under small perturbations of $\varepsilon$, resulting in an
  unstable~\cite{Ratschan:02b} constraint.}).  Furthermore we only deal with constraints
without negation symbols because one can easily eliminate negation symbols from quantified
constraints by pushing them down, and replacing atomic constraints of the form $\neg
(f\leq g)$ by $f>g$, and $\neg (f<g)$ by $f\geq g$, respectively.  For any quantified
constraint $\phi$, let $\SNeg{\phi}$ (the \emph{opposite of $\phi$}) be the quantified
constraints that results from $\neg\phi$ by eliminating the negation by pushing it down to
the predicates.

We require that every quantifier be bounded by an associated \emph{quantifier bound},
using expressions of the form $\exists x\STI I$ or $\forall x\STI I$, where $I$ is a
closed interval.

A \emph{variable assignment} is a function from the set of variables $V$ to $\RR$. We
denote the semantics of a constraint $\phi$, the set of variable assignments that make
$\phi$ true, by $\SSem{\phi}$. For example, $\SSem{x^2+y^2\leq 1}$ is the set of variable
assignments that assign values within the unit disc to $x$ and $y$.

For any variable assignment $d$, any variable $v\in V$ and any real number $r$, we
denote by $\SAssVar{d}{r}{v}$ the variable assignment that is equal to $d$ except that it
assigns $r$ to $v$.

Let $\SSInt$ be the set of closed real intervals. We denote by $I_1 \SIU I_2$ the smallest
interval containing both intervals $I_1$ and $I_2$.  A \emph{box assignment} is a set of
variable assignments that can be represented by functions from $V$ to $\SSInt$; that is,
it contains all the variable assignments that assign elements within a certain interval to
a variable. For example, for $V=\{ x, y \}$, the set of variable assignments that
assign an element of $[-1, 1]$ to both $x$ and $y$ is a box assignment---this set of
variable assignments can be represented by the function that assigns $[-1, 1]$ to both $x$
and $y$. 

From now on we will use a box assignment and its interval function representation
interchangeably.  For any box assignment $B$, any variable $v\in V$, and any interval $I$, we
denote by $\SAssVar{B}{I}{v}$ the box assignment that is equal to $B$ except that it assigns $I$
to $v$. In the context of closed constraints we denote by the Boolean value $\SFalse$ the empty
box assignment and by the Boolean value $\STrue$ the box assignment that assigns the set of real
numbers $\RR$ to each variable, and allow the usual Boolean operation on them. We denote by $\{
x\mapsto [-1, 1], y\mapsto\}$ a box assignment that assigns the interval $[-1, 1]$ to the
variable $x$ and an arbitrary interval to the variable $y$.

Traditionally, constraint programming
techniques~\cite{Benhamou:94a,Benhamou:97a,Sam-Haroud:96,Hentenryck:97} use boxes (i.e.,
Cartesian products of intervals) instead of box assignments. However, when working with
predicate logic, the additional flexibility of box assignments is very convenient in
dealing with the scoping of variables. For efficiency reasons, an actual implementation
might represent box assignments by boxes.





\section{A Framework for Reusable Pruning}
\label{sec:first-order-narr}

Remember that our approach will be to solve quantified constraints by a branch and prune
algorithm.  Fortunately, there is already a lot of work done on how to do pruning on
atomic constraints, and their conjunctions. The formal framework for this is the notion of
a narrowing operator~\cite{Benhamou:95,Benhamou:97a}, which specifies some properties
required of such an algorithm without regard to the concrete algorithm. In this paper we
generalize this notion to quantified constraints. This will allow us to reuse existing
theory, algorithms, and software implementing such narrowing operators. Readers who are
only interested in a concrete pruning algorithm and not in a formal framework for
reasoning about its properties can directly jump to Section~\ref{sec:narrowing-algorithm}.

The essential difference between our approach and the classical one is that quantified
constraints also store information about the range of variables within the constraint, and
so we allow a narrowing operator to modify constraints also.  For this we use pairs
$(\phi, B)$, where $\phi$ is a quantified constraint and $B$ is a box assignment. We call
such pairs \emph{bounded constraints}, and the second element of a bounded constraint its
\emph{free-variable bound}.

\begin{definition}
  A \emph{narrowing operator} is a function $N$ on bounded constraints such that for 
  bounded constraints $(\phi, B)$, and $(\phi', B')$, and for $N(\phi, B) = (\phi_N,
  B_N)$, and $N(\phi', B') = (\phi'_N, B'_N)$,

  \begin{itemize}
  \item $B \supseteq B_N$ (contractance), 
  \item $\SSem{\phi_N}\cap B_N = \SSem{\phi} \cap B_N$ (soundness),
  \item $B'\subseteq B$ implies $B'_N \subseteq B_N$ (monotonicity), and
  \item $N(N(\phi, B)) = N(\phi, B)$ (idempotence).
  \end{itemize}
\end{definition}

Note that we use a soundness condition here, instead of a
correctness condition: We require that the solution set of the resulting constraint be
the same only on the \emph{resulting} box, but not necessarily on the initial box. We will
ensure full correctness by the next definition.

Constraint programming techniques for continuous domains traditionally compute outer
approximations of the solution set. However, here we also want to compute inner
approximations for three reasons: First, we need to compute such inner approximations for
proving closed constraints to be true. Second, the solution set of quantified constraints
with inequality predicates usually does not consist of singular points, but of sets that
can have volume, and for many applications it is important to find points that are
guaranteed to be within this solution set. And third, available inner approximations can
speed up the computation of outer approximations and vice versa, because any element known
to be within the solution set, or known to be not in the solution set, does not need to be
inspected further. 

So we will allow two kinds of narrowing operators---one that only removes elements not
in the solution set, and one that only removes elements in the solution set.

\begin{definition}
  An \emph{outer narrowing operator} is a narrowing operator $\SNarC$ such that for every
  bounded constraint $(\phi, B)$, for the free-variable bound $B_{N}$ of $\SNarC(\phi,
  B)$, $B_{N}\supseteq B\cap\SSem{\phi}$.  An \emph{inner narrowing operator} is a narrowing
  operator $\SNarI$ such that for every bounded constraint $(\phi, B)$, for the
  free-variable bound $B_{N}$ of $\SNarI(\phi, B)$, $B_{N}\supseteq B\setminus\SSem{\phi}$\footnote{This definition of inner narrowing operator differs from the one used by
Benhamou and Goualard~\cite{Benhamou:00}.}.
\end{definition}


As discussed in the introduction, we get 
an inner narrowing operator from an outer narrowing operator by working on the opposite of the input:


\begin{theorem}
\label{thm:1}
  Let  $\SNarC$ be a function on bounded constraints and let $\SNarI(\phi, B):=
  (\SNeg{\phi_{N}}, B_{N})$ where $(\phi_{N}, B_{N})=\SNarC(\SNeg{\phi}, B)$. Then 
$\SNarC$ is an outer narrowing operator iff $\SNarI$ is an inner narrowing operator.
\end{theorem}

\begin{proof}
  Obviously $\SNarI$ is a narrowing operator iff $\SNarC$ is a narrowing
  operator. $\SNarC$ is outer narrowing iff $\SNarI$ is inner narrowing because for any
  bounded constraint $(\phi, B)$, $B\cap\SSem{\SNeg{\phi}} =
  B\setminus\SSem{\phi}$, and $B\setminus\SSem{\SNeg{\phi}} = B\cap\SSem{\phi}$. \qed
\end{proof}

A similar observation has already been used for the special case of quantified
constraints with one universal quantifier~\cite{Benhamou:00}. The above theorem allows
us to concentrate on outer narrowing operators from now on. We get the corresponding
inner narrowing operator for free by applying the outer narrowing operator on the opposite of
the input.


\section{Consistency of Quantified Constraints}
\label{sec:first-order-cons}

In constraint programming the notion of consistency is used to specify the pruning power
of narrowing operators.  In this section we generalize this approach to quantified
constraints. Again, readers who are only interested in a concrete algorithm and not in
formal reasoning about its properties, can skip this section.

Clearly, it is not possible to prune the empty set further. So we require the following from a
predicate on bounded constraints that we use for such specification purposes:

\begin{definition}
  A \emph{consistency property} is a predicate $C$ on bounded constraints such that for
  every constraint $\phi$, $C(\phi, \emptyset)$.
\end{definition}


\begin{example}
  For an atomic bounded constraint $(\phi, B)$, $BC(\phi, B)$ holds iff there is no
 (canonical-interval-wide) face
 of the hyperrectangle described by $B$ for
  which interval evaluation~\cite{Moore:66,Neumaier:90,Jaulin:01b} will prove that it
  contains no element of $\SSem{\phi}$. In this case we say that $\phi$ is
  \emph{box-consistent} wrt.  $B$~\cite{Benhamou:94a,Hentenryck:97}.
\end{example}

The strongest form of consistency achievable using floating-point numbers is:

\begin{example}
  For a bounded constraint $(\phi, B)$, $HC(\phi, B)$ holds iff there is no box
  assignment $B'$ with floating-point endpoints such that $B'\subset B$ and
  $\SSem{\phi}\cap B' = \SSem{\phi}\cap B$. In this case we say that $\phi$ is
  \emph{hull-consistent} wrt. $B$~\cite{Benhamou:97a}.
\end{example}

Note that in the constraint programming literature~\cite{Cleary:87,Benhamou:95,Benhamou:94a},
the definition of hull-consistency usually assumes, that the according constraints have been
decomposed into so-called \emph{primitive} constraints. We do \emph{not} follow this approach
here, because that would blur the borderline between consistency properties and symbolic
preprocessing of constraints.

The following is the strongest form of consistency that does not result in loss of
information, that is, for which an outer narrowing operator exists.

\begin{definition}
  For a bounded constraint $(\phi, B)$, $TC(\phi, B)$ holds iff there is no box
  assignment $B'$  such that $B'\subset B$ and
  $\SSem{\phi}\cap B' = \SSem{\phi}\cap B$. In this case we say that $\phi$ is
  \emph{tightly consistent} wrt. $B$.
\end{definition}

Now we can use consistency properties as specifications for the effectiveness of narrowing
operators:

\begin{definition}
  Given a consistency property $C$, a narrowing operator $\SNar$ \emph{ensures $C$} iff
  for all bounded constraints $(\phi, B)$, $C(\SNar(\phi, B))$ holds.
\end{definition}





Now we assume a certain consistency property on literals (i.e., atomic constraints and
their negations) and lift it to a corresponding consistency property on quantified
constraints.

\begin{definition}
\label{def:1}
Given a quantified constraint $\phi$ and a consistency property $C$ on literals,
let $\SLift{C}$ be the following predicate:

  \begin{itemize}
  \item if $\phi$ is a literal, then $\SLift{C}(\phi, B)$ iff $C(\phi, B)$
  \item if $\phi$ is of the form $\phi_{1}\wedge\phi_{2}$ then
     $\SLift{C}(\phi, B)$ iff 
      $\SLift{C}(\phi_{1}, B)$ and $\SLift{C}(\phi_{2}, B)$.
  \item if $\phi$ is of the form $\phi_{1}\vee\phi_{2}$ then 
     $\SLift{C}(\phi, B)$ iff $B = B_{1}\SIU B_{2}$ where 
      $\SLift{C}(\phi_{1}, B_{1})$ and $\SLift{C}(\phi_{2}, B_{2})$.
  \item if $\phi$ is of the form $Q y\STI I'\;\phi'$, where $Q$ is a quantifier, then 
    $\SLift{C}(\phi, B)$ iff $\SLift{C}(\phi', \SAssVar{B}{I'}{y})$.
 \end{itemize}
\end{definition}

\begin{theorem}
  For a consistency property $C$, $\SLift{C}$ is also a consistency property.
\end{theorem}

If for a bounded constraint $(\phi, B)$, $\SLift{C}(\phi, B)$ holds, we say that $(\phi, B)$
is \emph{structurally $C$-consistent}.

Note that, in the
above definition, recursion for quantification puts the quantifier bound into the
free-variable bound of the quantified constraint. This means that a narrowing operator
will also have to modify the quantifier bounds in order to achieve structural
consistency.

\begin{example}
\label{ex:3}
The bounded constraint $\bigl(\,\exists y\STI [0, 1] \bigl[ x^2+y^2\leq 1\wedge y\geq 0
\bigr]\,,\,\{ x\mapsto[-1,1], y\mapsto \}\,\bigr)$ is structurally tightly consistent, and it
will be the result of applying an according narrowing operator to an input such as
$\bigl(\,\exists y\STI [-2, 2] \bigl[ x^2+y^2\leq 1\wedge y\geq 0 \bigr]\,,\,\{ x\mapsto
[-2,2], y\mapsto\}\,\bigr)$.
\end{example}

Note that Definition~\ref{def:1} is compatible with the usual consistency notions for sets
of constraints~\cite{Benhamou:94a,Benhamou:99}. For example a set of atomic constraints
$\{ \phi_{1}, \dots, \phi_{n} \}$ is box-consistent wrt. a box assignment $B$ iff
$(\phi_{1}\wedge\dots\wedge\phi_{n}, B$) is $\SLift{BC}$-consistent.  In addition, the method
for solving constraints with one universally quantified variable by Benhamou and
Goualard~\cite{Benhamou:00} computes a special case of Definition~\ref{def:1}.



In the following sense, our definition of $\SLift{C}$-consistency is optimal (remember that
tight consistency is the strongest possible consistency property).

\begin{theorem}
\label{thm:2}
  A $\SLift{TC}$-consistent bounded constraint $(\phi, B)$, where $\phi$ contains neither
  conjunctions nor universal quantifiers, is $TC$-consistent.
\end{theorem}

\begin{proof}
  We proceed by induction over the structure of constraints. The atomic case trivially holds.
  Now assume constraints of the following types:

  \begin{itemize}

  \item For a $TC$-consistent bounded constraint of the form $(\phi_{1}\vee\phi_{2}, B)$,
    by definition, we have $B = B_{1}\SIU B_{2}$ where ${TC}(\phi_{1}, B_1)$ and
    ${TC}(\phi_{2}, B_2)$.  By the induction hypothesis both $(\phi_{1}, B_{1})$ and
    $(\phi_{2}, B_{2})$ are tightly consistent.  So for no box assignment $B'_{1}\subset
    B_{1}$, we have $B'_{1}\supseteq B_{1}\cap\SSem{\phi_{1}}$, and for no box assignment
    $B'_{2}\subset B_{2}$, we have $B'_{2}\supseteq B_{2}\cap\SSem{\phi_{2}}$. Thus also
    for no box assignment $B'\subset B_{1}\SIU B_{2}$, we have $B'\supseteq
    B\cap(\SSem{\phi_{1}}\cup\SSem{\phi_{2}})=B\cap\SSem{\phi_{1}\vee\phi_{2}}$.
  \item For a $\SLift{TC}$-consistent bounded constraint of the form $(\exists y\STI
    I'\;\phi', B)$, by definition $\SLift{TC}(\phi', \SAssVar{B}{I'}{y})$. Thus, by the
    induction hypothesis $(\phi', \SAssVar{B}{I'}{y})$ is tightly consistent. So for no box
    assignment $B_p' \subset \SAssVar{B}{I'}{y}$, $B_p'\supseteq
    \SAssVar{B}{I'}{y}\cap\SSem{\phi'}$. As a consequence also for no box assignment $B_p
    \subset B$, $B_p\supseteq B\cap\SSem{\exists y\STI I'\;\phi'}$, and so $(\exists y\STI
    I'\;\phi', B)$ is tightly consistent.

  \end{itemize} \qed

\end{proof}

\begin{figure}[htbp]
  \begin{center}
    \input{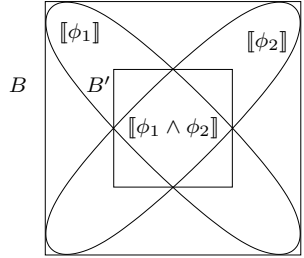}
    \caption{Conjunction---Structural Tight Consistency}
    \label{fig:dependency}
  \end{center}
\end{figure}

The fact that the above theorem does not hold for constraints with conjunctions is well
known~\cite{Benhamou:97a}.
It is illustrated in Figure~\ref{fig:dependency}, where both $\phi_{1}$ and $\phi_{2}$ are
tightly consistent wrt. the box $B$ (i.e. the larger box encloses the ellipses tightly), but
$\phi_{1}\wedge\phi_{2}$ is only tightly consistent wrt. the smaller box $B'$ (i.e., the
smaller, but not the larger box, encloses the intersection of the ellipses tightly).


\begin{figure}[htbp]
  \begin{center}
    \input{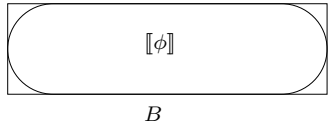}
    \caption{Universal Quantification---Structural Tight Consistency}
    \label{fig:univpruned}
  \end{center}
\end{figure}

For universal quantification there is a similar problem: In Figure~\ref{fig:univpruned},
$\phi$ is tightly consistent wrt. the box $B$, but when considering $\forall y\STI I\; \phi$
one can still narrow $B$ horizontally.  So any stronger consistency notion would have to
treat universal quantification differently from existential quantification.

\section{Pruning Algorithm}
\label{sec:narrowing-algorithm}


In this section we give an algorithm for pruning quantified constraints that can use an
arbitrary algorithm for pruning atomic constraints. This algorithm fulfills the formal
properties introduced in Sections~\ref{sec:first-order-narr}
and~\ref{sec:first-order-cons}.


The algorithm proceeds recursively according to the structure of constraints.  For
conjunctions this means the usual: We prune wrt. the individual sub-constraints, until
this does not result in any further improvements, that is, until we reach a fix-point. For
disjunctions we prune the individual sub-constraints and combine the result by taking the
smallest box assignment containing the union.

\begin{figure}[htbp]
   \centering    
   \input{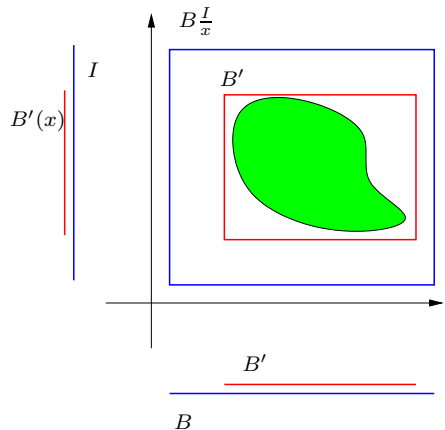}
   \caption{Existential Pruning}
   \label{fig:existPrune}
 \end{figure}

For existentially quantified bounded constraints of the form $(\exists x\STI I\;\phi, B)$
we proceed as shown in Figure~\ref{fig:existPrune}, where the horizontal axis represents
the free-variable bound $B$ (ignoring the component corresponding to the variable $x$) and
the vertical axis the quantifier bound $I$. Below and to the left of these axis we show
the changes on the corresponding elements.
We recursively prune the bounded sub-constraint $(\phi, \SAssVar{B}{I}{x})$ consisting of the
sub-constraint $\phi$, and the box assignment that is the same as $B$ except that it
assigns $I$ to the variable $x$. We use the result to remove these elements from the
free-variable bound $B$ and the quantifier bound $I$ for which recursive pruning showed
that they do not contain any solution. This results in the new free-variable bound $B'$
and quantifier bound $B'(x)$.



 For universally quantified bounded constraints of the form $(\forall x\STI I\;\phi, B)$,
 if pruning of the sub-constraints removes elements from $I$, the whole constraint is
 false, and we can replace $B$ by the empty set (Figure~\ref{fig:univPrune1}). If no such
 elements are removed then we just prune $B$ accordingly (Figure~\ref{fig:univPrune2}).


\begin{figure}[htbp]
  \centering
  \input{univPrune1.pstex_t}
  \caption{Universal Pruning}
  \label{fig:univPrune1}
\end{figure}

\begin{figure}[htbp]
  \centering
  \input{univPrune2.pstex_t}
  \caption{Universal Pruning}
  \label{fig:univPrune2}
\end{figure}


To formalize the above, we let $\SFix$ 
be a partial function such that, for a set of functions $\{ f_1,\dots,f_n \}$, for all
$a$, $\SFix(\{ f_1,\dots,f_n \})(a)$ is a fixpoint of applying $\{
f_1,\dots,f_n \}$ to $a$, if such a fixpoint exists, and is undefined, otherwise. Now we have:

\begin{definition} ~
\label{def:3}
  \begin{itemize}
  \item For atomic $\phi$, $\SNarC_{A}(\phi, B) = \SNarCP{A}(\phi, B)$, 
  \item $\SNarC_{A}(\phi_{1}\wedge\phi_{2}, B) = \SFix(\{ \SNarC'_{1}, \SNarC'_{2} \})(\phi_{1}\wedge\phi_{2},B)$\\
    \hspace*{0.5cm} where $\SNarC'_{1}(\phi_{1}\wedge\phi_{2}, B)= (\phi_{1}'\wedge \phi_{2}, B')$,
    where $(\phi_{1}', B')=\SNarC_{A}(\phi_{1}, B)$,\\ 
    \hspace*{0.5cm} and $\SNarC'_{2}(\phi_{1}\wedge\phi_{2}, B)= (\phi_{1}\wedge \phi_{2}', B')$,
    where $(\phi_{2}', B')=\SNarC_{A}(\phi_{2}, B)$.

  \item $\SNarC_{A}(\phi_{1}\vee\phi_{2}, B) = (\phi_{1}'\vee \phi_{2}', B_{1}'\SIU B_{2}')$ \\\hspace*{1.5cm}
    where $(\phi_{1}', B_{1}')=\SNarC_{A}(\phi_{1}, B)$ and $(\phi_{2}', B_{2}')=\SNarC_{A}(\phi_{2}, B)$

  \item $\SNarC_{A}(\exists x\STI I\;\phi, B) = (\exists x\STI B'(x)\;\phi', B')$, \\\hspace*{1.5cm}
    where $(\phi', B') = \SNarC_{A}(\phi, \SAssVar{B}{I}{x})$

  \item $\SNarC_{A}(\forall x\STI I\;\phi, B)$ = $(\forall x\STI I\;\phi', D)$, \\\hspace*{1.5cm}
    where $(\phi', B') = \SNarC_{A}(\phi, \SAssVar{B}{I}{x})$\\\hspace*{1.5cm}
    and $d\in D$ iff for all $r\in I$, $\SAssVar{d}{r}{x}\in B'$
  \end{itemize}
\end{definition}


\begin{example}
  For the input $\bigl(\,\exists y\STI [-2, 2] \bigl[ x^2+y^2\leq 1\wedge y\geq 0
  \bigr]\,,\,\{ x\mapsto[-2,2], y\mapsto\}\,\bigr)$ already used in Example~\ref{ex:3}, a
  narrowing operator based on tight consistency applies itself recursively
  to $(x^2+y^2\leq 1\wedge y\geq 0\,,\,\{ x\mapsto [-2,2], y\mapsto [-2,2]\})$.
  Repeated applications of the atomic narrowing operator---until a fixpoint is
  reached---will create the constraint $(x^2+y^2\leq 1\wedge y\geq 0\,,\,\{
  x\mapsto[-1,1], y\mapsto [0,1]\})$. As a final result we get $\bigl(\,\exists y\STI [0,
  1] \bigl[ x^2+y^2\leq 1\wedge y\geq 0 \bigr]\,,\,\{ x\mapsto [-1,1],
  y\mapsto\}\,\bigr)$.
\end{example}

\begin{example}
  For the input $(\forall x\STI [-2, 2]\, x\geq 0, \{ x\mapsto \})$, the algorithm will first narrow
  $(x\geq 0, \{ x\mapsto [-2, 2] \})$ to $(x\geq 0, \{ x\mapsto [0, 2]\})$ and then create $(\forall x\STI [-2, 2]\, 
  x\geq 0, \emptyset)$, indicating that the constraint is false.
\end{example}

In the rest of this section we will use the results of Section~\ref{sec:first-order-narr}
and~\ref{sec:first-order-cons} for studying the properties of the introduced pruning
algorithm. Readers not interested in these formal properties can directly jump to
Section~\ref{sec:solver} for an according solver.

Note that the fixed-point operator $\SFix$ could result in a partial function, that is,
the algorithm could fail to terminate.  For ensuring termination, we require that
pruning atomic constraints eventually terminates even when intermingled with
shrinking of the free-variable bound by other operations:

\begin{definition}
\label{def:2}
A narrowing operator $N$ is \emph{finitely contracting} iff there is no infinite sequence
$(\phi_{1}, B_{1}), (\phi_{2}, B_{2}), \dots$ for which for all $k\in\NN$, for $(\phi',
B')=N(\phi_{k}, B_{k})$, $\phi_{k+1}=\phi'$ and $B_{k+1}$ is a strict subset of $B'$.
\end{definition}

This property usually holds for practical implementations, because of the finiteness of
floating point numbers.

\begin{lemma}
  If $\SNarCP{A}$ is a finitely contracting narrowing operator then $\SNarC_{A}$ is a total
  function.
\end{lemma}

\begin{proof}
  We assume that $\SNarCP{A}$ is finitely contracting but $\SNarC_{A}$ is not total.  This
  can only happen if $\SFix(\{ \SNarC'_{1}, \SNarC'_{2} \})(\phi_{1}\wedge\phi_{2},B)$ is
  undefined. Consider the sequence $(\phi^{1}_{1}\wedge\phi^{1}_{2}, B_{1}),
  (\phi^{2}_{1}\wedge\phi^{2}_{2}, B_{2}), \dots$ of bounded constraints created by
  repeated applications of $\SNarC'_{1}$ and $\SNarC'_{2}$. Here $(\phi^{1}_{1}, B_{1}), (\phi^{2}_{1},
  B_{2}), \dots$ is an infinite sequence as in Definition~\ref{def:2}. So $\SNarC_{A}$ is not
  finitely contracting, and by induction also $\SNarCP{A}$ is not finitely contracting---a
  contradiction.  \qed
\end{proof}


The algorithm fulfills the properties needed by the formal framework of narrowing operators and
gives a unique result (despite the non-unique definition of fixpoint operator): 

\begin{theorem}
  For every finitely contracting (atomic) outer narrowing operator $\SNarCP{A}$, $\SNarC_{A}$ is
  a unique outer narrowing operator.
\end{theorem}

\begin{proof}
  Contractance and idempotence hold by easy induction. For proving that $\SNarC_{A}$ is
  outer narrowing, sound and monotonic we proceed by induction. The ground case of atomic
  constraints holds by definition. Now we have:

  \begin{itemize}
  \item Obviously the composition of two narrowing operators is also a narrowing operator.
    So, for constraints of the form $\phi_{1}\wedge\phi_{2}$ we just need to show that
    both $\SNarC'_{1}$ and $\SNarC'_{2}$ are outer narrowing. For $(\phi_{1}',
    B_{1}')=\SNarC(\phi_{1}, B)$ and $(\phi_{2}', B_{2}')=\SNarC(\phi_{2}, B)$, by the
    induction hypothesis $B_{1}'\supseteq B\cap\SSem{\phi_{1}}$ and $B_{2}'\supseteq
    B\cap\SSem{\phi_{2}}$.  Thus also $B_{1}'\cap B_{2}'\supseteq
    B\cap\SSem{\phi_{1}}\cap\SSem{\phi_{2}}=B\cap\SSem{\phi_{1}\wedge\phi_{2}}$. The
    induction step for soundness and monotonicity is easy.
  \item For constraints of the form $\phi_{1}\vee\phi_{2}$, for $(\phi_{1}',
    B_{1}')=\SNarC(\phi_{1}, B)$ and $(\phi_{2}', B_{2}')=\SNarC(\phi_{2}, B)$, by the
    induction hypothesis $B_{1}'\supseteq B\cap\SSem{\phi_{1}}$ and $B_{2}'\supseteq
    B\cap\SSem{\phi_{2}}$. Thus also 
    $B_{1}'\SIU B_{2}'\supseteq
     B_{1}'\cup B_{2}'\supseteq
     B\cap(\SSem{\phi_{1}}\cup\SSem{\phi_{2}})=
     B\cap\SSem{\phi_{1}\vee\phi_{2}}$. The induction step for soundness and monotonicity
    is easy.
  \item For constraints of the form $\exists x\in I\; \phi$, the induction step for
    outer narrowing is easy. For soundness we have to prove that $\SSem{\exists x\in I\;\phi}
    \cap B' = \SSem{\exists x\in B'(x)\;\phi'}\cap B'$, where $B' = \SNarC_{A}(\phi,
    \SAssVar{B}{I}{x})$. Now by the outer narrowing property $B'\supseteq B\cap\SSem{\phi}$,
    and so $\SSem{\exists x\in I\;\phi} \cap B' = \SSem{\exists x\in B'(x)\;\phi}\cap B'$.
    This is equal to $\SSem{\exists x\in B'(x)\;\phi'}\cap B$, because by the induction
    hypothesis $\SSem{\phi'}\cap B' = \SSem{\phi} \cap B'$.
  \item For constraints of the form $\forall x\in I\; \phi$, for outer narrowing we have to
    prove that $D\supseteq B\cap \SSem{\forall x\in I\; \phi}$, where $D$ is defined as in
    the corresponding rule of Definition~\ref{def:3}. So we assume a variable assignment $d$
    that is both in $B$ and $\SSem{\forall x\in I\; \phi}$, and prove that $d\in D$. This
    means that we have to prove that for all $r\in I$, $\SAssVar{d}{r}{x}\in B'$, where
    $(\phi', B') = \SNarC_{A}(\phi, \SAssVar{B}{I}{x})$. This is clearly the case by the
    semantics of universal quantification and the induction hypothesis.
    
    For soundness we have to prove that $\SSem{\forall x\in I\; \phi} \cap D =
    \SSem{\forall x\in I\; \phi'}\cap D$, where $D$ is as above.  By the quantifier
    semantics it suffices to prove that $\SSem{\phi}\cap \SAssVar{D}{I}{x}=
    \SSem{\phi'}\cap \SAssVar{D}{I}{x}$.  This holds, because for all variable assignments
    $d\in D$, for all $r\in I$, $\SAssVar{d}{r}{x}\in B'$, and moreover, by the induction
    hypothesis $\SSem{\phi'}\cap B' = \SSem{\phi} \cap B'$.

  \end{itemize}

  The uniqueness of the fixpoint operator follows from contractance and monotonicity of
  narrowing operators~\cite{Cousot:77a,Apt:99a}.
 \qed
\end{proof}

By easy induction we also get: 

\begin{theorem}
  $\SNarC_{A}$ ensures $\SLift{A}$-consistency.
\end{theorem}

By applying Theorem~\ref{thm:1} we get a corresponding inner narrowing operator $\SNarI_{A}$ from
$\SNarC_{A}$. Note, however, that $\SNarC_{A}$ and $\SNarI_{A}$ do not commute, and
$\SNarC_{A}\circ\SNarI_{A}$ is not idempotent.

As for the classical conjunctive case, the complexity of the algorithm in a floating-point
implementation is polynomial in the problem dimension (the number of floating point
numbers that one can remove from the quantification bounds is polynomial). So, as
desired, pruning is efficient compared to expensive exhaustive search, and even more so
compared to the doubly exponential complexity of symbolic
solvers~\cite{Collins:75,Caviness:98}. 

Note however, that the cardinality of the usual float-point representations is so high
that, in addition, one should take care that the worst-case complexity is not reached in practice.

\section{Solver}
\label{sec:solver}

Now a branch-and-prune algorithm can do pruning according to Definition~\ref{def:3}, and
branching means replacing sub-constraints of the form $\forall x\STI I\;\phi$ by $\forall
x\STI I_{1}\;\phi\;\wedge\;\forall x\STI I_{2}\;\phi$, or sub-constraints of the form
$\exists x\STI I\;\phi$ by $\exists x\STI I_{1}\;\phi\;\vee\;\exists x\STI I_{2}\;\phi$,
where $I=I_{1}\cup I_{2}$. We assume branching to
be \emph{fair}, in the sense that every bound will eventually be split (finding such a
strategy is easy, but finding an optimal branching strategy is highly non-trivial).

For disproving a closed constraint $\phi$ we repeatedly branch and prune the input
constraint $(\phi, \STrue)$ until $\STrue$ is pruned to $\SFalse$ (remember that $\SFalse$
is an abbreviation for the empty box assignment). For proving we do the same on the
opposite of $\phi$. For computing the truth-value we do both things in parallel.  

A solver for open quantified constraints could for example use the following
specification:

\begin{description}
\item[Given: ] ~
  \begin{itemize}
  \item A quantified constraint $\phi$ with $n$ free variables,
  \item $B\subseteq\RR^{n}$,
  \item $\varepsilon\in\RR^{+}$
  \end{itemize}
\item[Find: ] Sets of boxes $Y$, $N$ s.t. 
  \begin{itemize}
  \item all elements of $Y$ are in the solution set of $\phi$,
  \item all elements of $N$ are not in the solution set of $\phi$,
  \item Vol($B\setminus \bigcup Y\setminus \bigcup N$)$\leq\varepsilon$
  \end{itemize}
\end{description}

This specification allows the user to decide on the trade-off between run-time and
precision. When choosing a large $\varepsilon$, only a small part of the solution set of $\phi$
within $B$ will be characterized by $Y$ and $N$, when choosing an $\varepsilon$ close to zero,
almost the whole set will be characterized.

An according solver is an easy extension of the closed case that would record the boxes
that narrowing of the input constraint proved to be not in the solution set, and narrowing
of the opposite of the input constraint proved to be in the solution set.  Furthermore,
in addition to branching the quantified variables it also has to branch the free-variable
bound. We call the resulting algorithm, a \emph{parallel branch-and-prune solver}.



For discussing termination of such a branch-and-prune solver it is important to see, that
the problem of computing truth-values is undecidable. So it is impossible to find an
algorithm that terminates always. A solution to this problem is to require termination for
all, except very special cases. 

For this we observe that the truth-value/solution set of a quantified constraint can be
numerically unstable~\cite{Ratschan:02b}. An example is the quantified constraint $\exists
x\STI [-1, 1]\; -\!x^2\geq 0$ which is true, but becomes false under arbitrarily small
positive perturbations of the constant $0$.  As a consequence, it is not possible to
design an algorithm based on approximation that will always terminate (with a correct
result). Note that this situation is similar for most computational problems of continuous
mathematics (e.g., solving linear equations, solving differential equations). However, as
in these cases, most inputs are still numerically stable (in fact, in a certain, realistic
model this is the case with probability one~\cite{Ratschan:01d}). One can even argue that,
philosophically speaking, the stable problems are exactly the problems that model
real-life problems in a meaningful way.
  
It is beyond the scope of this paper to present all the formal details for characterizing
stable quantified constraints and we will introduce the necessary concepts in a
semi-formal way. For this we replace the discrete notion of truth of a quantified
constraint by a continuous notion~\cite{Ratschan:02b}. We interpret universal quantifiers
as infimum operators, existential quantifiers as supremum operators, conjunction as
minimum, disjunction as maximum, atomic constraints of the form $f>g$ or $f\geq g$ as the
function denoted by $f-g$, and atomic constraints of the form $f<g$ or $f\leq$ g as the
function denoted by $g-f$.  We call the result the \emph{degree of truth} of a quantified
constraint and denote it by $\SSemD{\phi}$ for any constraint $\phi$. This function
assigns to every variable assignment a real value that is independent of the variables
that are not free in $\phi$.  The idea is that the degree of truth is greater or equal
zero for variable assignments that make $\phi$ true, and less or equal zero for variable
assignments that make $\phi$ false.  One can prove~\cite{Ratschan:02b} that the problem of
computing the truth value of a closed quantified constraint is numerically stable (or:
\emph{stable}) iff its degree of truth is non-zero.
  
  We assume that the given narrowing operator for atomic constraints eventually succeeds for
  stable inputs:
  
\begin{definition}
  An outer narrowing operator $\SNarCP{A}$ is \emph{converging} iff for all atomic
  constraints $\phi$ and sequences $B^{0}\supseteq B^{1}\supseteq\dots$ such that 
  \begin{itemize}
  \item for all $i\in\NN$, $B^{i}\supseteq B^{i+1}$,
  \item and $\bigcap_{i\in\NN} B^{i}= \{ d \}$, where the degree of truth of $\phi$ at the
    variable assignment $d$ is negative,
  \end{itemize}
 there is a $k$, such that for all $l\geq
  k$, the free-variable bound of $\SNarCP{A}(\phi, B^{l})$ is empty.
\end{definition}


Note that this trivially holds for tight consistency. For hull and box consistency it is
necessary that the number of bits used in floating point computation is high enough for a given
sequence of boxes. For box consistency, in addition, interval evaluation has to
converge for all atomic constraints. This is the case if they only contain continuous functions
such as $+$, $\times$, $\exp$, and $\sin$, on which we can implement interval evaluation, see
Theorem~2.2~\cite{Jaulin:01b}).

However, the above property is in general impossible to fulfill for any narrowing operator
based on fixed-precision floating-point arithmetic. Still, one can easily overcome this
difficulty, by using sufficient precision~\cite{Revol:02} (see Theorem 2.1.5~\cite{Neumaier:90}).
Moreover, the application of our method to real-life problems has shown that
usual double-precision floating-point arithmetic almost always suffices in practice.

\begin{lemma}
\label{lem:6}
  Let $\SNarCP{A}$ be  a  converging outer narrowing operator 
    and let the sequence $(\phi^{1}, B^{1}), (\phi^{2}, B^{2}), \dots$ be such that
      \begin{itemize}
      \item for all $i\in\NN$, $B^{i}\supseteq B^{i+1}$, and
      \item         $\bigcap_{i\in\NN} B^{i}= \{ d \}$ such that the degree of truth of
        $\phi$ at the variable assignment $d$ is negative, 
      \item for all $i\in N$, $\phi^{i+1}$ results from $\phi^{i}$ by branching, and
      \item for all $\varepsilon>0$ there is a $k$ such that for all $l\geq k$, the volume of all quantification
        sets in $\phi^{l}$ is less or equal $\varepsilon$.\footnote{This item formalizes the notion of a fair
branching strategy.}
      \end{itemize}
   Then there is a $k$ such that for all $l\geq k$, the free-variable bound of $\SNarC_{A}(\phi^{l},
      B^{l})$ is empty.
\end{lemma}

\begin{proof}
  We proceed by induction over the structure of the constraint $\phi^{1}$.
   For atomic constraints the lemma holds because $\SNarCP{A}$ is converging.  Now consider
  the following cases:

  \begin{itemize}
  \item For constraints of the form $\phi_{1}\wedge\phi_{2}$,
    $\SSemD{\phi_{1}\wedge\phi_{2}}(d)=\min\{\SSemD{\phi_{1}}(d),\SSemD{\phi_{2}}(d)\}$ 
    being negative implies that either $\SSemD{\phi_{1}}(d)$ is negative or
    $\SSemD{\phi_{2}}(d)$ is negative. 
    Therefore at least one of the sequences $(\phi_{1}^{1}, B^{1}), (\phi_{1}^{2}, B^{2}),
    \dots$ and $(\phi_{2}^{1}, B^{1}), (\phi_{2}^{2}, B^{2}), \dots$ where $\phi_{1}^{i}$
    is the sub-constraint of $\phi^{i}$ corresponding to $\phi_{1}$ and $\phi_{2}^{i}$ is the
    sub-constraint of $\phi^{i}$ corresponding to $\phi_{2}$, fulfills the preconditions of
    the induction hypothesis. Let $r\in\{
    1, 2 \}$ be the number of this sequence. Then there is a $k$, such that for all $k\geq l$, the
    free-variable bound of $\SNarC_{A}(\phi_{r}^{l}, B^{l})$ is empty. Thus, by definition of
    $\SNarC_{A}$, the corresponding free-variable bound is also empty in the original sequence.

  \item 
    For constraints of the form $\phi_{1}\vee\phi_{2}$,
    $\SSemD{\phi_{1}\vee\phi_{2}}(d)=\max\{\SSemD{\phi_{1}}(d),\SSemD{\phi_{2}}(d)\}$ 
    being negative implies that both $\SSemD{\phi_{1}}(d)$ and
    $\SSemD{\phi_{2}}(d)$ are negative.
    Therefore both sequences $(\phi_{1}^{1}, B^{1}), (\phi_{1}^{2}, B^{2}), \dots$ and
    $(\phi_{2}^{1}, B^{1}), (\phi_{2}^{2}, B^{2}), \dots$ where $\phi_{1}^{i}$ is the
    sub-constraint of $\phi^{i}$ corresponding to $\phi_{1}$ and  $\phi_{2}^{i}$ is the
    sub-constraint of $\phi^{i}$ corresponding to $\phi_{2}$,
    fulfill the
    preconditions of the induction hypothesis. As a consequence there is a $k_{1}$,
    such that for all $l\geq k_{1}$, the free-variable bound of $\SNarC_{A}(\phi_{1}^{l}, B^{l})$ is empty,
    and there is a $k_{2}$,
    such that for all $l\geq k_{2}$, the free-variable bound of $\SNarC_{A}(\phi_{1}^{l}, B^{l})$ is empty.
     Thus, by definition of $\SNarC_{A}$, for all $l\geq \max\{k_1, k_2\}$ the 
     free-variable bound of the $l$-th element in the original sequence is empty.
     
   \item Constraints of the form $\forall x\in I\; \phi'$, are replaced by branching into
     the form $\forall x\in I_{1}\; \phi' \;\wedge\;\dots\;\wedge\; \forall x\in I_{k}
     \phi'$.  Since the degree of truth of $\forall x\in I\; \phi'$ at $d$ is negative, by
     definition of infimum, there is a $b\in I$ for which the degree of truth of $\phi'$
     at $d\times b$ is negative.  Consider the sequence for which the $i$-th element
     consists of the branch of $\phi^{i}$ that contains $d\times b$, and of $B^{i}$. This
     sequence fulfills the preconditions of the induction hypothesis, and as a consequence
     there is a $k$, such that for all $k\geq l$, the $k$-th free-variable bound in this
     sequence is empty.  Thus, by definition of $\SNarC_{A}$, the corresponding
     free-variable bound is also empty in the original sequence.
     
   \item Constraints of the form $\exists x\in I\; \phi'$, are replaced by branching into
     the form $\exists x\in I_{1}\; \phi' \;\vee\;\dots\;\vee\; \exists x\in I_{k} \phi'$.
     Since the degree of truth of $\exists x\in I\; \phi'$ at $d$ is negative, by
     definition of supremum, for all $b\in I$ the degree of truth of $\phi'$ at $d\times
     b$ is negative.  This means that each sequence for which the $i$-th element consists
     of a branch of $\phi^{i}$ and of $B^{i}$ fulfills the preconditions of the induction
     hypothesis, and as a consequence there is a $k$, such that for all $k\geq l$, the $k$-th
     free-variable bound in this sequence is empty.  Thus, by definition of $\SNarC_{A}$,
     the corresponding free-variable bound is also empty in the original sequence.

\qed
  \end{itemize} 
\end{proof}

This implies:

\begin{lemma}
\label{lem:4}
  A branch-and-prune algorithm for disproving a closed constraint succeeds iff the degree
  of truth of the input is negative.
\end{lemma}

From Lemma~\ref{lem:4} and its dual version we get:

\begin{theorem}
  For stable inputs a parallel branch-and-prune solver eventually computes the truth value
  for closed inputs, and fulfills the above solver specification for open inputs.
\end{theorem}


\section{Efficient Implementation}
\label{sec:effic-impl}

In this section we show how to extend the basic solver for allowing an efficient
implementation.

\subsection{Connectives with Arbitrary Arity}

 


The first step for arriving at an efficient implementation, is to treat conjunctions and
disjunctions not as binary operators, but as operators with arbitrary arity (see
Definition~\ref{def:3}). It is easy to adapt the according algorithms and proofs.

\subsection{Quantifier Blocks}

Treat quantifiers of the same kind in blocks. That is, quantify over a whole vector of
variables at once, using quantifier bounds that are boxes instead of intervals.  This
allows more flexibility for branching.

\subsection{Removing Empty Disjunctive Branches}

Pruning might show that one of the branches of a disjunction has an empty solution
set. Currently (see Definition~\ref{def:3}) this information is forgotten. In order to
prevent this we simply remove the corresponding sub-constraint in such as case. 

\subsection{Combination with Negated Constraint}

The parallel branch-and-prune solver developed in Section~\ref{sec:solver}
independently works on proving and disproving the input constraint. For proving, it
employs a branch-and-prune procedure on the negation of the input, for disproving, on the
input itself.  Working on the input and its negation separately has the disadvantage that
information computed for one is not used for the other. In order to improve this, we do
both on the same constraint, repeatedly negating it in between.  The result is
Algorithm~\ref{alg:4} for closed constraints and Algorithm~\ref{alg:3} for open
constraints. Here ``Branch'' and ``Prune'' do the obvious, except that in the second
algorithm ``Branch'' takes a bounded constraint and returns a set of bounded constraints
that either
\begin{itemize}
\item contains two elements whose union of bounds is equal to the input bound, or 
\item contains one element with a quantifier split into a conjunction (in the case of a
  universal quantifier), or a disjunction (in the case of an existential quantifier).
\end{itemize}

\begin{algorithm}
\caption{Combined Solver for Closed Constraints}
\label{alg:4}
\begin{algorithmic}
\REQUIRE a closed quantified constraint $\phi$
\ENSURE the truth-value of $\phi$
\STATE $\mbox{unknown}\leftarrow\STrue$
\WHILE{$\mbox{unknown}$}
  \STATE $\mbox{neg}\leftarrow\STrue$
  \STATE $\phi'\leftarrow\SFalse$
  \STATE{$(\SNeg{\phi}, \mbox{unknown})\leftarrow\mbox{Prune}(\SNeg{\phi}, \mbox{unknown})$}
  \WHILE{$\mbox{unknown}$ and $\phi\neq\phi'$}
    \STATE $\phi'\leftarrow\phi$
    \STATE $\mbox{neg}\leftarrow\;\mbox{not neg}$
    \IF{neg}
      \STATE{$(\SNeg{\phi}, \mbox{unknown})\leftarrow\mbox{Prune}(\SNeg{\phi}, \mbox{unknown})$}
    \ELSE
      \STATE{$(\phi, \mbox{unknown})\leftarrow\mbox{Prune}(\phi, \mbox{unknown})$}
    \ENDIF
  \ENDWHILE
  \IF{$\mbox{unknown}$}
    \STATE{$\phi\leftarrow\mbox{Branch}(\phi)$}
  \ENDIF
\ENDWHILE
\STATE \textbf{return} neg
\end{algorithmic}
\end{algorithm}

\begin{theorem}
  Algorithms~\ref{alg:4} and~\ref{alg:3} are correct and terminate for stable inputs.
\end{theorem}

\begin{proof}
  We just show the proof for Algorithm~\ref{alg:4}, the other case is similar.

  If neg is false at the return statement, then pruning succeeded for $\phi$, and so
  it is false. Otherwise, neg is true and so pruning succeeded for $\SNeg{\phi}$
  and $\phi$ is true. 
  
  The innermost while loop always terminates because there are only finitely many floating
  point numbers, and so the narrowing operator can only do finitely many changes after
  which $\phi=\phi'$. Termination of the overall loop again is a consequence of
  Lemma~\ref{lem:6}.  \qed
\end{proof}

\begin{algorithm}
\caption{Combined Solver for Open Constraints}
\label{alg:3}
\begin{algorithmic}
\REQUIRE a quantified constraint $\phi$, a box $B$, and a positive real number $\varepsilon$
\ENSURE $Y$, a list of boxes on which $\phi$ is true, and $N$, a list of boxes on which $\phi$
is false, such that the volume of $B\setminus\bigcup Y\setminus\bigcup N$ is less than $\varepsilon$.
\STATE $U\leftarrow \{ (\phi, B) \}$
\STATE $Y\leftarrow\emptyset$
\STATE $N\leftarrow\emptyset$
\WHILE{$\mbox{Vol}(\bigcup U)\geq\varepsilon$}
  \STATE choose and remove a bounded constraint $(\phi_U, B_U)$ from $U$
  \STATE $\mbox{neg}\leftarrow\STrue$
  \STATE $(\phi_U', B_U')\leftarrow(\phi_U, B_U)$
  \STATE{$(\SNeg{\phi_U}, B_U)\leftarrow\mbox{Prune}(\SNeg{\phi_U}, B_U)$}
  \STATE $Y \leftarrow Y\cup B_U'\setminus B_U$
  \WHILE{$B_U$ is non-empty and $(\phi_U, B_U)\neq(\phi_U', B_U')$}
    \STATE $(\phi_U', B_U')\leftarrow (\phi_U, B_U)$
    \STATE $\mbox{neg}\leftarrow\;\mbox{not neg}$
    \IF{neg}
      \STATE{$(\SNeg{\phi_U}, B_U)\leftarrow\mbox{Prune}(\SNeg{\phi_U}, B_U)$}
      \STATE $Y \leftarrow Y\cup B_U'\setminus B_U$
    \ELSE
      \STATE{$(\phi_U, B_U)\leftarrow\mbox{Prune}(\phi_U, B_U)$}
      \STATE $N \leftarrow N\cup B_U'\setminus B_U$
    \ENDIF
  \ENDWHILE
  \IF{$\mbox{Vol}(\bigcup U)\geq\varepsilon$}
    \STATE $U\leftarrow U\cup \mbox{Branch}(\phi, B_U)$ 
  \ENDIF
\ENDWHILE
\end{algorithmic}
\end{algorithm}

\subsection{Branching Strategy}

For arriving at an implementation, a good strategy for choosing a (free-variable or
quantification) bound for branching is crucial. We did not yet try to arrive at a
theoretically well-founded or even optimal strategy. However, the following approach seems
to work well in practice:

In Algorithm~\ref{alg:3}, for every element of the set $U$ we store the level of the last
splitting done (viewing constraints as trees), and for each conjunction or disjunction (of
the original constraint, not the ones created by branching) the last branched
sub-constraint. We choose the bounded constraint $(\phi_U, B_U)$ with largest $B_U$ and 
branch a sub-constraint
\begin{itemize}
\item that is one level below the last splitting, or (if this is impossible) on the
  highest level,
\item with the maximum volume of the quantifier bound for conjunctions or disjunctions
  created by branching, which is
\item the next sub-constraint for all other conjunctions and disjunctions.
\end{itemize}

\subsection{Incremental Disjunctive Pruning}

The disjunctive case in Definition~\ref{def:3}, has the disadvantage that it stores the
intermediate results of narrowing all sub-constraints until computing the box union $\SIU$
of all of them. We can avoid this by using an incremental algorithm instead that
intermingles the recursive calls with taking the box union of the result.




\subsection{Shortcut for Disjunctions}
\label{sec:shortc-disj}

When doing incremental pruning of disjunctions we might detect that the result will be the
input box, before inspecting all sub-constraints. We can leave the according loop already
at this point.

\subsection{Reusing Dual Information}
\label{sec:reusing-information}


Sometimes an atomic narrowing operator computes the information that
narrowing the negation will fail. For example, assume that a bound
$[-2, 2]$ of a univariate atomic constraint has been pruned to $[-1,
1]$. For many narrowing operators this implies that the constraint
does not hold on $-1$ and $1$. Therefore, we cannot use the opposite
constraint to prune $[-1, 1]$ further.

In the case of box consistency, which we use in our implementation, one works on atomic
constraints one variable after the other. For each atomic sub-constraint/variable pair we
can get the information that pruning or pruning on the opposite will not succeed, that is
that this sub-constraint or its opposite is consistent.  We store this information by
assigning to each atomic sub-constraint two sets of variables (the \emph{mark} and the
\emph{opposite mark}). Furthermore, in order to reflect the situation for box consistency
we assume that consistency also takes into account variables.




\begin{definition}
\label{def:5}
We call a bounded constraint $(\phi, B)$ \emph{correctly marked}
iff for every atomic sub-constraint $\phi'$ of $\phi$, every box assignment $B'$ that
results from $B$ by replacing all variables of $B$ bounded by $\phi$ by these bounds, and
every variable $v$ in the mark of $\phi'$, $(\phi', B', v)$ is consistent.
\end{definition}

Sometimes we change the bound of a variable. In this case some of the marks might not stay
valid. For example, take a constraint of the form $\exists x\in I_x\exists y\in I_y\;
[\phi_1\wedge \phi_2]$, where $\phi_1$ is an atomic constraint that contains both
variables $x$ and $y$, but $\phi_2$ is an atomic constraint that only contains $y$.  After
pruning, the marks and opposite marks of both $\phi_1$ and $\phi_2$ can be set to $\{ x,
y\}$, indicating that no further pruning is possible. After branching the quantifier of
$y$ we have to remove the marks of both copies of $\phi_1$, but we can keep the marks of
the copies of $\phi_2$ since $\phi_2$ contains no variable affected by the branching.
Therefore, further pruning will know that calls to the atomic narrowing operator of $\phi_2$
are not necessary.

So denote by $\SNotify{\phi, V}$ the result of replacing all marks of sub-constraints that
contain variables in $V$ by the empty set.

\begin{lemma}
\label{lem:1}
For every bounded constraint $(\phi, B)$ that is correctly, and for every box
$B'$ such that for all $v\not\in V$, $B'(v)=B(v)$, $\SNotify{\phi, V}, B'$ is correctly marked.
\end{lemma}

\begin{proof}
  All sub-constraints of $\SNotify{\phi, V}$ that contain $V$ have empty marks and
  Definition~\ref{def:5} requires nothing from them. For all bounded sub-constraints that
  do not contain $V$, the corresponding bound inherited from $B$ is the same as the one
  from $B'$, and therefore one does not have to change the corresponding marks. \qed
\end{proof}

\begin{definition}
  A narrowing operator \emph{preserves marks} iff after applying it to a bounded
  constraint that is correctly marked the result is again correctly marked.
\end{definition}

We assume that we have an atomic narrowing operator that preserves marks and opposite
marks. Of course this can be easily done by always setting the marks to the empty set.
But, in our implementation, we will try to set them as large as possible.

Now, whenever applying the narrowing operator on atomic constraint/variable pairs we check
the marks before. For more complicated constraints we have to adapt the narrowing operator
of Definition~\ref{def:3} such that it updates the marks accordingly.  This means that for
disjunctions, if the box union is different from the boxes resulting from narrowing an
individual constraints, then we have to do notification on it.  In a similar way, for
conjunctions, when computing the fixpoint, every time narrowing succeeds for a
sub-constraint, we have to do notification for all other sub-constraints.  Clearly, by
Lemma~\ref{lem:1} the adapted narrowing operator preserves marks and opposite marks.

Also for branching we have to do according notification for the changed variables.  When
using adapted branching and pruning in Algorithms~\ref{alg:4} and \ref{alg:3}, we set all
marks to the empty set at the beginning and preserve marks and opposite marks
throughout.

Clearly the resulting solver does less calls to the narrowing operator for projection
constraints than the original one. Furthermore this also gives an improvement for the case
of unquantified constraints for which the solution set does not consist of finitely many,
isolated solutions.


\section{Timings}
\label{sec:timings}

We have implemented the algorithm described in this paper using as the atomic narrowing operator
an algorithm~\cite{Hong:94b} that computes something similar to box consistency. In
Table~\ref{tab:1} we compare this implementation described in this paper with an implementation
of the older algorithm using ``cylindrical box decomposition''~\cite{Ratschan:01a} which also
used the same atomic narrowing operator.  Here the heading ``0.1 old'' refers to running the
older algorithm until it leaves not more than a fraction of $1/10$ of the solution space
unknown, the heading ``first'' refers to running the current algorithm till it finds the first
true box, and the heading ``0.1'' refers to running the current algorithm in the same way as the
older one.

Columns headed by
``Time'' list the time in seconds needed to solve the problem where $\varepsilon$ means
less than a second and $\infty$ more than $10$ minutes; columns headed by ``Hits'' list
the number of calls to the atomic narrowing operator for the new algorithms (this is a
good efficiency measure because it ignores implementation details, and because atomic
narrowing takes the largest part of the overall runtime); and columns headed by ``Boxes''
list the total number of boxes created (true, false, and unknown boxes).

Examples starting with McCallum are from computational
geometry~\cite{McCallum:93}---asking the questions whether there exists a solution to a
given system of inequalities. The example ``circle'' simply computes a description of the
unit disc $x^2+y^2\leq 1$, ``silaghi'' is a system of inequalities describing the geometry
of a simple mechanical problem~\cite{Silaghi:01}, the example
``anderson2'' is a system of polynomial inequalities from control engineering~
\cite{Anderson:75,Abdallah:96,Garloff:99}, the example ``anderson2\_proj'' computes the
projection of the former into two-dimensional space, the example ``termination'' proves
the termination of a certain term-rewrite system~\cite{Collins:91}. All examples starting
with ``robust'' are taken from robust
control~\cite{Dorato:00,Fiorio:93,Malan:97,Dorato:95,Jaulin:96,Jaulin:02} and laid out on
a web-site~\cite{Ratschan:01e}.

For all examples we push quantifiers inside as much as possible by transforming $\forall
(\phi_1\wedge\phi_2)$ to $(\forall\phi_1)\wedge(\forall\phi_2)$ and the dual for
existential quantification. Furthermore, in the few cases, where no quantification bounds
were available, we introduced new, very large ones.

\begin{table}[htbp]
  \centering
\begin{tabular}[htb]{|l|rr|rrr|rrr|} \hline
        & \multicolumn{2}{c|}{0.1 old} & \multicolumn{3}{c|}{first} & \multicolumn{3}{c|}{0.1}\\ 
Example & Time & Boxes & Time & Hits & Boxes       & Time & Hits & Boxes \\ \hline 
McCallum\_2\_1 & $\varepsilon$ &  & $\varepsilon$ & 22 &  & $\varepsilon$ & 22 &  \\
McCallum\_2\_2 & $\varepsilon$ &  & $\varepsilon$ & 157 &  & $\varepsilon$ & 157 &  \\
McCallum\_2\_3 & $\varepsilon$ &  & $\varepsilon$ & 2854 &  & $\varepsilon$ & 2854 &  \\
McCallum\_2\_4 & $\varepsilon$ &  & $\varepsilon$ & 153 &  & $\varepsilon$ & 153 &  \\
anderson2 & $\varepsilon$ & 391 & 10.95 & 150343 & 575 & $\varepsilon$ & 3372 & 250 \\
anderson2\_proj & 16.81 & 3466 & $\varepsilon$ & 1129 & 21 & 2.69 & 21058 & 620 \\
circle & $\varepsilon$ & 42 & $\varepsilon$ & 19 & 10 & $\varepsilon$ & 60 & 26 \\
robust-1 & $\varepsilon$ & 23 & $\varepsilon$ & 13 & 3 & $\varepsilon$ & 121 & 16 \\
robust-2 & $\infty$ &  & 90.74 & 663011 & 1266 & $\infty$ & $\infty$ &  \\
robust-3 & $\infty$ &  & $\varepsilon$ & 344 & 5 & $\varepsilon$ & 936 & 9  \\
robust-5 & $\infty$ & & $\varepsilon$ & 376 & 11 & $\infty$ & $\infty$ &  \\
robust-6 & $5.4$ & 112 & $\varepsilon$ & 34 & 3 & 256.14 & 564326 &  4309 \\
silaghi1 & $\varepsilon$  & 19 & $\varepsilon$ & 25 & 6 & $\varepsilon$ & 25 & 6 \\
termination & $\infty$ & &  $\varepsilon$ & 90 & & $\varepsilon$ & 90 & 
\end{tabular}
\caption{Comparison with Cylindrical Box Decomposition}
\label{tab:1}
\end{table}

As one can see, except for the example ``robust-6'' the number of generated boxes is much
smaller for the new algorithm. For the examples where it is possible to make a clear
comparison of the run-times, the new algorithm is also faster (again with the exception of
``robust-6''). An analysis of the behavior for the outlier ``robust-6'' shows that in
this case our branching heuristics do not work very well---alternative heuristics show a
much better behavior. This suggests that a detailed study of such heuristics---expanding
results for a simpler branch-and-bound approach~\cite{Ratschan:02c} can still result in
large improvements of the method.



In Table~\ref{tab:2} we show the results of some of the algorithm improvements introduced
in Section~\ref{sec:effic-impl}.  We chose these that need a non-trivial implementation
effort or these for which it is not totally clear that they improve the efficiency of the
algorithm. These are the ones described in Section~\ref{sec:shortc-disj} and
Section~\ref{sec:reusing-information}.

\begin{table}[htbp]
   \centering  
\begin{tabular}[htb]{|l|rr|rr|rr|rr|} \hline
         & \multicolumn{2}{c|}{no reuse/no sh.cut} & \multicolumn{2}{c|}{no reuse/sh.cut} &
 \multicolumn{2}{c|}{reuse/no sh.cut} & \multicolumn{2}{c|}{reuse/sh.cut} \\
 Example & Time & Hits & Time & Hits & Time & Hits & Time & Hits \\ \hline
McCallum\_2\_1 & $\varepsilon$ & 22 & $\varepsilon$ & 22 & $\varepsilon$ & 22 & $\varepsilon$ & 22 \\
McCallum\_2\_2 & $\varepsilon$ & 182 & $\varepsilon$ & 169 & $\varepsilon$ & 150 & $\varepsilon$ & 157 \\
McCallum\_2\_3 & $\varepsilon$ & 10059 & $\varepsilon$ & 3005 & $\varepsilon$ & 3760 & $\varepsilon$ & 2854 \\
McCallum\_2\_4 & $\varepsilon$ & 163 & $\varepsilon$ & 165 & $\varepsilon$ & 138 & $\varepsilon$ & 153 \\
anderson2 & $\varepsilon$ & 4034 & $\varepsilon$ & 4251 & $\varepsilon$ & 2964 & $\varepsilon$ & 3372 \\
anderson2\_proj & 8.49 & 78985 & 8.54 & 65401 & 7.57 & 55443 & 7.54 & 55244 \\
circle & $\varepsilon$ & 95 & $\varepsilon$ & 95 & $\varepsilon$ & 60 & $\varepsilon$ & 60 \\
robust-1 & $\varepsilon$ & 138 & $\varepsilon$ & 139 & $\varepsilon$ & 116 & $\varepsilon$ & 121 \\
robust-2 & $\infty$ & $\infty$ & $\infty$ & $\infty$ & $\infty$ & $\infty$ & $\infty$ & $\infty$ \\
robust-3 & $\varepsilon$ & 1623 & $\varepsilon$ & 1115 & $\varepsilon$ & 1003 & $\varepsilon$ & 936 \\
robust-5 & $\infty$ & $\infty$ & $\infty$ & $\infty$ & $\infty$ & $\infty$ & $\infty$ & $\infty$ \\ 
robust-6 & 147.89 & 637216 & 285.52 & 702734 & 69.8 & 255163 & 256.14 & 564326 \\
silaghi1 & $\varepsilon$ & 31 & $\varepsilon$ & 27 & $\varepsilon$ & 27 & $\varepsilon$ & 25 \\
termination & $\varepsilon$ & 282 & $\varepsilon$ & 102 & $\varepsilon$ & 156 & $\varepsilon$ & 90 
\end{tabular}
\caption{Comparison of Improvements}
\label{tab:2}
\end{table}

One can conclude that reusing dual information (Section~\ref{sec:reusing-information})
always improves the algorithm, sometimes significantly.  This phenomenon also occurs for
examples that do not contain any quantifiers. On the other hand, for taking
shortcuts for disjunctions (Section~\ref{sec:shortc-disj}), the influence on efficiency is
inconclusive, especially when combined with the former improvement. 

Note that often one can get even better run-times by symbolically eliminating linearly
quantified variables before~\cite{Weispfenning:88,Loos:93}.  Unfortunately, in some cases
the result can be very large, destroying the positive effect of the eliminated variable.
Future work will investigate this behavior in detail.

\section{Relation to Classical Algorithms}
\label{sec:relat-class-algor}

A.~Tarski~\cite{Tarski:51} showed that quantified constraints over the reals with equality
and inequality predicates, multiplication and addition admit quantifier elimination.
Adding additional function symbols (e.g., $\sin$, $\tan$), usually removes this
property~\cite{Richardson:68,Dries:88,MacIntyre:96}. Using the method in this paper one
can still compute useful information for these cases, provided that the input is
numerically stable.

The complexity bound supplied by Tarski's method has been improved several
times~\cite{Collins:75,Renegar:92,Basu:94}---but the problem is inherently doubly
exponential~\cite{Davenport:88,Weispfenning:88} in the number of quantifier alternations, and
exponential in the number of variables.

The only general algorithm for which a practically useful implementation exists, is the
method of \emph{quantifier elimination by cylindrical algebraic
  decomposition}~\cite{Collins:75}.  This algorithm employs similar branching as the
algorithm presented in this paper.  However, its branching operation is much more
complicated because it branches into a finite set of \emph{truth-invariant cells}, that
is, into pieces whose value can be computed by evaluation on a single \emph{sample point}.
For being able to do this, its quantifier bounds can depend on the free variables, and
branching is done based on information from \emph{projection polynomials}. For
implementing these operations one needs expensive real algebraic number computations.

Instead of branching, quantifier elimination by \emph{partial cylindrical algebraic
  decomposition}~\cite{Collins:91} employs pruning in a similar sense as described in this
paper. However it still decomposes into truth-invariant cells, which again needs expensive
computation of projection polynomials, and real algebraic numbers.

In contrast to this, the narrowing operator provided in this paper is cheap, and can do
pruning in polynomial time. As a result, we have a clear separation between polynomial
time pruning, and exponential branching. So we have a way of working around the high
worst-case complexity of the problem, whenever a small amount of branching is necessary.

For inputs with free variables all of these algorithms produce symbolic output that is
equivalent to the input, but quantifier-free. This output can be huge. In many
applications, such output is only considered a transformation of the problem, but not a
solution. In contrast to this, our algorithm produces explicit numerical output, that one
can directly visualize for dimensions less than three.


\section{Conclusion}
\label{sec:conclusion}

In this paper we have provided an algorithm for solving quantified inequality constraints
over the reals. Although this is an undecidable problem, the algorithm terminates for all,
except pathological (i.e., unstable) inputs.

The result has several advantages over earlier approaches: Compared to symbolic
approaches~\cite{Collins:75,Caviness:98} it is not restricted to polynomials, and avoids
complicated and inefficient computation with real algebraic numbers. Furthermore it
decreases the necessity for expensive space decomposition by extracting information using
fast consistency techniques. Compared to earlier interval approaches that could deal with
quantifiers of some form, it provably terminates for all except unstable inputs, and can
either handle a more general case~\cite{Jaulin:96,Malan:97,Benhamou:00,Shary:99}, or
provides a much cleaner, more elegant, and efficient framework~\cite{Ratschan:01a}.

As a side-effect, this paper even improves the current methods for (unquantified)
numerical constraint satisfaction problems in the case where the solution set does not
consist of finitely many, isolated solutions.

In future work we will explore optimal branching
strategies~\cite{Ratschan:02c,Kreinovich:01,Csendes:97}, exploit continuity information
for efficiently dealing with equalities~\cite{Ratschan:02d}, exploit the structure of
quantified constraints in special problem domains, and provide an implementation that
allows the flexible exchange of different atomic narrowing operators.

\ifthenelse{\boolean{acm}}{
\begin{acks}}{}
This work has been supported by 
a Marie Curie fellowship of the European Union under contract number HPMF-CT-2001-01255.
\ifthenelse{\boolean{acm}}{
\end{acks}}{}

\ifthenelse{\boolean{acm}}{
\bibliographystyle{ACM/acmtrans}
}{
\bibliographystyle{abbrv}
}

\bibliography{sratscha,sratscha_own}

\ifthenelse{\boolean{acm}}{
\begin{received}
Received November 2002;
revised January 2004 and January 2005;
accepted January 2005
\end{received}
}{}

\end{document}